\documentstyle[pra,aps,epsf,twocolumn]{revtex}  
\begin{document}                
\title{Radiative atom-atom 
interactions in optically dense media: Quantum corrections to the
Lorentz-Lorenz formula}
\author{Michael Fleischhauer and Susanne F.~Yelin}
\address{Sektion Physik, Ludwig-Maximilians Universit\"at M\"unchen,
D-80333 M\"unchen, Germany}
\address{and: Dept. of Physics, Texas A \& M University, College Station
TX 77843-4242, U.S.A.}
\date{\today}
\maketitle 


\begin{abstract} 
Generalized single-atom Maxwell-Bloch equations for  
optically dense media are derived taking into account 
non-cooperative radiative atom-atom interactions. 
Applying a Gaussian approximation and formally eliminating 
the degrees of freedom of the quantized
radiation field and of all but a probe atom leads to
an effective time-evolution operator for the probe atom. 
The mean coherent amplitude of the local field 
seen by the atom is shown to be given 
by the classical Lorentz-Lorenz relation.
The second-order correlations of the field
lead to terms that describe relaxation or pump processes and 
level shifts due to multiple scattering or reabsorption of 
spontaneously emitted photons. In the Markov limit a nonlinear
and nonlocal single-atom density matrix equation is derived.
To illustrate the effects of the quantum corrections we discuss
amplified spontaneous emission and radiation trapping in a 
dense ensemble of initially inverted two-level atoms and 
the effects of radiative interactions on 
intrinsic optical bistability in coherently driven systems.
\end{abstract}





\section{Introduction}


The interaction of the radiation field with 
a dilute ensemble of atoms 
is usually described in the semiclassical and dipole approximation 
by the well-known Maxwell-Bloch equations. This description fails to 
be accurate, however, when a dense medium is considered. 

Since the early work of H.~A.~Lorentz and L.~Lorenz \cite{LL}
it is known that the classical local field, that couples to an 
atom in a dense medium, differs from the macroscopic (Maxwell) 
field by a term proportional to the medium polarization \cite{LL2,Bowden93}.  
The most prominent effects of the Lorentz-Lorenz (LL)-correction
in dense media are the change of the linear index of refraction 
according to the Clausius-Mossotti relation \cite{LL2}, the
enhancement of nonlinear susceptibilitiess \cite{Bloembergen65}, 
shifts and deformation of resonance lines \cite{Boyd,Woerdman},
intrinsic optical bistability \cite{intr_bist,Rand}, and piezo-photonic
switching \cite{piezo}.

On the other hand
the quantum nature of the radiative atom-atom interaction can drastically
influence the behavior of the ensemble. In the extreme case of 
anisotropic, high-density samples, excited atoms can cooperatively emit
spontaneous photons, a phenomenon known as superradiance  
\cite{Dicke,Haroche,Andreev}. But even if the system does not fulfill the
conditions for cooperative evolution, the presence of spontaneous photons 
and the associated effects like amplified spontaneous emission 
(or superluminescence) and radiation trapping 
\cite{Holstein} can not be neglected. Imprisonment of incoherent photons 
especially affects otherwise long-lived ground-state coherences. We therefore
expect radiative atom-atom interactions to be important in areas such as
resonant linear and nonlinear optics based on atomic phase coherence 
\cite{HI,NLO}, cooling of atoms and Bose-Einstein condensation via
velocity-selective coherent population trapping \cite{VSCPT} and optical
computing. 

Another important effect of large atomic densities is the increase of atomic
collisions. Here we will  not consider these effects, 
however, and focus our attention entirely on radiative interactions.

In the present paper we study the atomic evolution in 
a dense medium irradiated by external coherent light fields. 
The macroscopic classical radiation field in the medium obeys
Maxwell's equations with the mean atomic polarization as source term.
To derive equations of motion for the many-atom system, 
 we start from a nonrelativistic quantized interaction 
Hamiltonian. Thus interactions between the atoms mediated by the
quantized radiation field such as reabsorption and scattering
of spontaneous photons are taken into account. 

Our aim is to derive an effective single-atom density-matrix equation.
For this we introduce an interaction picture with the radiation field
coupling to all other atoms. Assuming a Gaussian 
(and therefore classical) statistics of the
interacting field,
we can formally eliminate its degrees of freedom from the probe-atom
time evolution. In the Markov limit of short-lived field
correlations this yields a density-matrix equation for the probe atom. 
We will show that the mean coherent amplitude seen
by the probe atom differs from the 
macroscopic Maxwell field by a term proportional to the mean polarization
of the medium in agreement with 
the classical Lorentz-Lorenz relation \cite{LL2}.
In addition, the density matrix equation contains
relaxation and level-shift terms, which describe 
reabsorbing and multiple scattering of spontaneously emitted photons. 
The corresponding relaxation rates and frequency shifts are proportional 
to the spectrum of the incoherent part of the radiation inside the medium.
This spectrum is also the Fourier-transform of a certain 2-time 
Greensfunction, for which we derive a Dyson equation. A formal
solution of the Dyson equation allows to express the incoherent spectrum
in terms of atomic variables. 
Thus we eventually obtain a closed, {\it nonlinear}
and spatially {\it nonlocal} density matrix equation of Lindblad-type.

Our paper is organized as follows. In Sec.~II we derive the effective
single-particle time-evolution operator by formally eliminating
the degrees of freedom of the quantized radiation field 
interacting with the background atoms. In Gaussian approximation 
this operator contains first and second-order field cumulants.
In the Markov limit of spectrally broad field correlations,
a density matrix equation is obtained. 
In Sec.~III we show that the 
first-order term leads
to the Lorentz-Lorenz relation between the coherent amplitude of the 
local field, the mean field amplitude in the medium (Maxwell field), and the
mean polarization. 
In Sec.~IV we derive a Dyson equation for the second-order field cumulants or
2-point Greensfunctions and formally solve them in terms of single-atom
density matrix elements. The resulting nonlinear density matrix equation
is discussed in Sec.V 
for the examples of amplified spontaneous emission and radiation
trapping in an inhomogeneously broadened system of initially excited two-level
atoms and intrinsic optical bistability in a strongly driven
dense ensemble of two-level atoms.


\section{Effective time-evolution of atoms}


\subsection{Formal elimination of the quantized radiation field}


We here consider an ensemble of atoms interacting with the
quantized radiation field under conditions which justify the  dipole 
and rotating-wave approximation (RWA). Since we are interested in the
dynamics of a single atom, we distinguish a probe atom at position
${\vec r}_0$ with a dipole operator $\vec p$
and environment atoms at positions ${\vec r}_j$
whose dipole operators are denoted by ${\vec d}^j$. 
The Hamiltonian of the system is given by
\begin{eqnarray}
H&=&\sum_j H_0^j + H_{\rm field}
- \vec p\cdot \left[{\vec { E}}(\vec r_0) +{\vec {\cal E}}(\vec
r_0)\right]\nonumber\\  
&& - \sum_{j\ne 0}\, {\vec d\,}^j\cdot
\left[{\vec { E}}(\vec r_j) +{\vec{\cal E}}(\vec r_j)\right],
\end{eqnarray}
where $H_0^j$ and $H_{\rm field}$ are the free Hamiltonians 
of the $j$th atom and the quantized radiation field respectively, and
we have split the field in an operator component
$E$ and an external classical driving 
field $ {\cal E}$.
We use an interaction
picture where the time evolution is described by
\begin{eqnarray}
S&=&T\exp\left\{-\frac{i}{\hbar}\int_{-\infty}^\infty\! \!\! d{\tau}\enspace 
V_{\rm p}(\tau)\right\}\;=\nonumber \\
&&T\exp\left\{\frac{i}{\hbar}\int_{-\infty}^\infty\!\! d\tau
\enspace 
p(\tau) \left[ E({\vec r}_0,\tau)+{\cal E}({\vec r}_0,\tau)\right]
\right\},\label{S}
\end{eqnarray}
were $T$ denotes time-ordering and the
field operator $E$ still  contains
the coupling to all other atoms. For notational simplicity we have suppressed
vector indices of the dipole moment and electric field.
 With the help of  (\ref{S}) any 
(time-ordered) correlation function of  probe-atom operators $A_H$ 
and $B_H$ in the
Heisenberg-picture (subscript ''$H$'') can be related to interaction 
picture operators via
\begin{eqnarray}
&&\langle T^{-1}[A_H(t_1) A_H(t_2)] T[B_H(t_3)B_H(t_4)]\rangle =\label{corr}\\
&&\quad\quad\left\langle T^{-1} [S^{-1} A(t_1)A(t_2)]
T [S B(t_3)B(t_4)]\right\rangle,\nonumber
\end{eqnarray}
where $\langle\cdots\rangle$ stands for $Tr\{\rho_0\cdots\}$ with $\rho_0=
\rho(-\infty)$ being the initial density operator at $t=-\infty$. 

A very helpful formal simplification of Eq.(\ref{corr}) can be achieved
by introducing the so-called Schwinger-Keldysh time contour $C$ \cite{Keldysh}
shown in Fig.~1
which starts at $t=-\infty$, goes to $t=+\infty$ and back $t=-\infty$. Each
 physical time correspond two times on the contour. 
A time ordering operator $T_C$ is introduced, which is 
 identical to $T$ on the upper branch $(+)$
and to $T^{-1}$ on the lower branch $(-)$ of the contour and orders all
operators with time arguments on $(-)$  to the left of 
those with time arguments on $(+)$.

\begin{figure}
\epsfxsize=7cm
\centerline{\leavevmode\epsffile{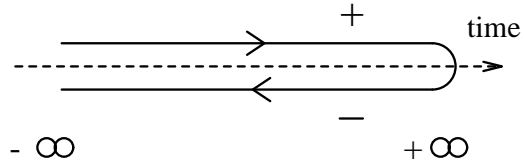}}
\caption{Schwinger-Keldysh time contour}
\end{figure}

 With these definitions 
we can write Eq.(\ref{corr}) with a single exponential time-evolution
operator. This will considerably simplify the following elimination
procedure.  
\begin{eqnarray}
&&\langle[T^{-1} A_H(t_1) A_H(t_2)][T B_H(t_3)B_H(t_4)]\rangle = \\
&&\quad\quad\langle T_C [S_C A(t_1^{-})A(t_2^{-}) B(t_3^{+}) B(t_4^{+})]
\rangle,\nonumber
\end{eqnarray}
where the superscripts $\pm$ specify the branch of the contour, and
\begin{equation}
S_C=T_C\exp\left\{-\frac{i}{\hbar}\int_C \!\! d{\check\tau}\enspace 
V_{\rm p}({\check\tau})\right\},
\end{equation}
with ${\check\tau}$ denoting a time on $C$.

We now formally eliminate 
the degrees of freedom of the
quantized radiation field and the environment atoms 
by tracing over the corresponding states.  
In order to express the expectation value of an exponential operator
again as an exponential operator, i.e. as a new --  effective -- 
time-evolution operator, we use a generalization of the
cumulant generating function for a classical
stochastic variable $X$ \cite{Gardiner,MFl}:
\begin{equation}
\Bigl\langle\exp\{sX\}\Bigr\rangle_X=\exp\left\{\sum_{m=0}^\infty 
\frac{s^m}{m!}\langle\langle X^m\rangle\rangle\right\},\label{cum}
\end{equation}
where the $\langle\langle X^m\rangle\rangle$ are the cumulants, which
 have the following explicit form
\begin{eqnarray}
\langle\langle X\rangle\rangle &=& \langle X\rangle, \\
\langle\langle XY\rangle\rangle &=& \langle XY\rangle 
-\langle X\rangle\langle Y\rangle,\quad{\rm etc.}
\end{eqnarray}
As can bee seen from (\ref{cum}), the elimination procedure leads in general
to an infinite number of terms in the effective action. 
To make the problem
tractable, we will however assume that the radiation field 
is Gaussian, i.e. that all cumulants
$\langle\langle E^m\rangle\rangle$ with $m > 2$ vanish
identically. This is a consistent and for our purposes  well justified
approximation. With this we find 
\begin{eqnarray}
\lefteqn{S_C^{\rm eff}=\left\langle S_C\right\rangle_{\rm field}
\label{Seff} \;}\\
&&\nonumber  T_C\exp\Biggl\{\frac{i}{\hbar}\int_C\!\! d{\check\tau}
\, p({\check \tau})
\Bigl[{\cal E}({\vec r}_0,{\check \tau})
+\langle E({\vec r}_0,{\check \tau})\rangle
\Bigr]\\
&&-\frac{1}{2\hbar^2}\int_C\!\! d{\check\tau}_1\int_C \!\! 
d{\check \tau}_2\, p({\check \tau}_1)\, {\cal D}({\vec r}_0,{\check \tau}_1;
{\vec r}_0,{\check \tau}_2)\, p({\check \tau}_2)
\Biggr\},\nonumber
\end{eqnarray}
where 
\begin{equation}
{\cal D}_{\mu\nu}({\check 1},{\check 2})
=\Bigl\langle \Bigl\langle  T_C\, E_\mu(\vec r_1,{\check\tau_1}) 
E_\nu(\vec r_2,{\check\tau_2}) \Bigr\rangle\Bigr\rangle\label{GF1}
\end{equation}
is a (tensorial) Greensfunction (GF) of the interacting electric field,
and  we have used the abbreviations ${\check 1}\equiv \vec r_1,\tau_1$ and 
${\check 2}\equiv\vec r_2,\tau_2$.
Note, that we used a
short notation, and  $p(\check 1)\, {\cal D}(\check 1,\check 2)\, p(\check 2)$ 
in Eq.(\ref{Seff}) 
should read $\sum_{\alpha,\beta=1}^3\, 
p_\alpha(\check 1)\,{\cal D}_{\alpha\beta}(\check 1,\check 2)\, 
p_\beta(\check 1,\check 2)$. 
We now apply the rotating-wave approximation. For this we introduce
slowly varying positive and negative frequency components,
\begin{eqnarray}
p(\check \tau) &=& p^+(\check \tau)+p^-(\check\tau) =
{\widetilde p\, }^{ +}(\check \tau) e^{-i\omega \tau}
 +
{\widetilde p\,}^{-}(\check \tau) e^{i\omega \tau},\\
E(\check \tau) &=& E^+(\check \tau)+E^-(\check \tau)
={\widetilde E}^{+}(\check \tau) e^{-i\omega \tau} +
{\widetilde E}^{-}(\check \tau) e^{i\omega \tau},
\end{eqnarray}
with $\omega$ being the transition frequency of the considered probe atom, 
and neglect
combinations of the type $p^{+} E^{+}$ and $p^{-} E^{-}$.
Thus we have
\begin{eqnarray}
\lefteqn{S_C^{\rm eff} \;=}\nonumber\\
&& T_C\exp\Biggl\{\frac{i}{\hbar}\int_C\!\! d{\check\tau}
\Bigl[ p^+({\check \tau})
{\cal E}_L^-({\vec r}_0,{\check \tau})+
p^-({\check \tau}){\cal E}_L^+({\vec r}_0,{\check \tau})\Bigr]\nonumber\\
&&\qquad\quad -\frac{1}{2\hbar^2}\int_C\!\! d{\check\tau}_1\int_C \!\! 
d{\check \tau}_2\Bigl[ {\widetilde p\,}^+({\check \tau}_1)\, 
D({\vec r}_0,{\check \tau}_1;
{\vec r}_0,{\check \tau}_2)\,  {\widetilde p\,}^-({\check \tau}_2)
\nonumber \\
&&\qquad\qquad\qquad +  {\widetilde p\,}^-({\check \tau}_1)\, 
C({\vec r}_0,{\check \tau}_1;
{\vec r}_0,{\check \tau}_2)\,  {\widetilde p\,}^+({\check \tau}_2)\Bigr]
\Biggr\},\label{Seff_RWA}
\end{eqnarray}
where 
\begin{equation}
{\cal E}_{L\mu}(\vec r,t)=
{\cal E}_{\mu}(\vec r,t)+\langle E_{\mu}(\vec r,t)\rangle
\end{equation}
is the local field seen by the probe atom, and
\begin{eqnarray}
\lefteqn{ D_{\mu\nu}({\vec r}_0,{\check \tau}_1;{\vec r}_0,{\check \tau}_2)
\;=}\nonumber\\
&& \langle\langle T_C E_\mu^-({\vec r}_0,{\check \tau}_1)
 E_\nu^+({\vec r}_0,{\check \tau}_2)\rangle\rangle\, 
e^{-i\omega(\tau_1-\tau_2)}
 ,\\
\lefteqn{C_{\mu\nu}({\vec r}_0,{\check \tau}_1;{\vec r}_0,{\check
\tau}_2)\;=} \nonumber \\
&& \langle\langle T_C E_\mu^+({\vec r}_0,{\check \tau}_1)
 E_\nu^-({\vec r}_0,{\check \tau}_2)\rangle\rangle\, 
e^{+i\omega(\tau_1-\tau_2)}.
\end{eqnarray}


\subsection{Markov approximation and single-atom density matrix equation}


The effective single-atom time-evolution operator (\ref{Seff_RWA})
leads in general to integro-differential equations of motion. We therefore 
restrict the discussion to situations that justify 
a Markov approximation, i.e. we assume
that the characteristic decay time of 
field cumulants is short compared to
the characteristic time of the atomic dynamics. This is the case, for example,
if the medium is inhomogeneously broadened or under
quasi-stationary conditions. We note that the Markov 
approximation used in the present paper rules out superradiance,
since the superradiance time is shorter than the decay time 
of field correlations \cite{Haroche,Andreev}.
In order to describe fast {\it cooperative} processes, pair-correlations
between different atoms need to be taken into account as a dynamical 
variable. This will be discussed in a future publication, where we
derive an effective density matrix equation for atom pairs\cite{super}.

The Markov approximation amounts to assuming a $\delta$-correlation
of $D_{\mu\nu}$ and $C_{\mu\nu}$ in  {\it physical} times.
\begin{eqnarray}
D_{\mu\nu}^{AB}(\tau,\tau^\prime) &=& D_{\mu\nu}^{AB}(\tau)\, 
\delta(\tau-\tau^\prime),\\
C_{\mu\nu}^{AB}(\tau,\tau^\prime) &=& C_{\mu\nu}^{AB}(\tau)\, 
\delta(\tau-\tau^\prime),
\end{eqnarray}
with $A,B\in \{+,-\}$ explicitly denoting the contour branches.
It is convenient to introduce dimensionless dipole operators 
$\sigma,\sigma^\dagger$, such that
$p_\mu^+(t)=\wp_\mu\, \sigma_\mu(t)$ and $p_\mu^-(t)=\wp_\mu\, 
\sigma^\dagger_\mu(t)$ (and corresponding relations for the slowly-varying
variables).
With this we eventually arrive at
\begin{eqnarray}
\lefteqn{S_C^{\rm eff} T_C\exp\Biggl\{}\nonumber \\
&& \frac{i\wp_\mu}{\hbar}\int_{-\infty}^\infty\!\!\! d{\tau}
\Bigl[ 
\sigma_\mu({\tau}_+)
{\cal E}_{L\mu}^-({\vec r}_0,\tau)
- \sigma_\mu({\tau}_-)
{\cal E}_{L\mu}^-({\vec r}_0,\tau)\nonumber\\
&&\qquad\qquad
+\sigma_\mu^\dagger({\tau}_+)
{\cal E}_{L\mu}^+({\vec r}_0,\tau)
- \sigma_\mu^\dagger({\tau}_-)
{\cal E}_{L\mu}^+({\vec r}_0,\tau)\Bigr]\nonumber\\
&&-\int_{-\infty}^\infty\!\!\! d\tau\, \frac{\Gamma_{\mu\nu}(\omega,\tau)}{2}\,
\Bigl[\sigma_\mu(\tau_+)\sigma_\nu^\dagger(\tau_+)+ \nonumber \\
&&\qquad\qquad\sigma_\mu(\tau_-)\sigma_\nu^\dagger(\tau_-)-2\sigma_\mu(\tau_-)
\sigma_\nu^\dagger(\tau_+)\Bigr]\label{Seff_RWA2}\\
&&-\int_{-\infty}^\infty\!\!\! d\tau\, \left(\frac{
\Gamma_{\mu\nu}(\omega,\tau)}{2}+
\frac{\gamma_{\mu\nu}(\omega,\tau)}{2}\right)\,
\Bigl[\sigma_\nu^\dagger(\tau_+)\sigma_\mu(\tau_+)+ \nonumber \\ 
&& \qquad\sigma_\nu^\dagger(\tau_-)\sigma_\mu(\tau_-) -
2\sigma_\nu^\dagger(\tau_-) 
\sigma_\mu(\tau_+)\Bigr]\nonumber\\
&&+\frac{i}{\hbar}\int_{-\infty}^\infty\!\!\! d\tau\, 
 H_{\mu\nu}(\omega,\tau)\,
\Bigl[\sigma_\mu(\tau_+)\sigma_\nu^\dagger(\tau_+) - \nonumber \\
&&\qquad \sigma_\mu(\tau_-)\sigma_\nu^\dagger(\tau_-)
-\sigma_\nu^\dagger(\tau_+)\sigma_\mu(\tau_+)+
\sigma_\nu^\dagger(\tau_-)\sigma_\mu(\tau_-)\Bigr]\nonumber\\
&&+\frac{i}{\hbar}
\int_{-\infty}^\infty\!\!\! d\tau\, h_{\mu\nu}(\omega,\tau)\,
\Bigl[
\sigma_\nu^\dagger(\tau_+)\sigma_\mu(\tau_+)-
\sigma_\nu^\dagger(\tau_-)\sigma_\mu(\tau_-)\Bigr]\Biggr\}.\nonumber
\end{eqnarray}
\narrowtext
The lower indices $\pm$ at the time argument denote the branch on the
Schwinger-Keldysh contour which is relevant for operator ordering
under the action of $T_C$.
The first term in (\ref{Seff_RWA2}) 
describes the interaction of the probe atom with the
local field in RWA.
\begin{eqnarray}
\lefteqn{\Gamma_{\mu\nu}(\omega,t)\;=} \nonumber \\
&&\frac{\wp_\mu\wp_\nu}{\hbar^2}\,
\int_{-\infty}^\infty\!\!\!d\tau\, \langle\langle E^-_\mu(\vec r_0,t)
E^+_\nu(\vec r_0,t+\tau)\rangle\rangle\, e^{i\omega\tau}\label{G_1}\\
&=&\frac{\wp_\mu\wp_\nu}{\hbar^2}\, 
{\widetilde D}_{\mu\nu}^{-+}(\vec r_0,\omega;t)\nonumber 
\end{eqnarray}
is a positive hermitian matrix, whose eigenvalues describe decay and
pump rates induced by the incoherent photons
 inside the medium. 
Eq.(\ref{G_1}) has a simple physical interpretation.
The incoherent radiation inside the medium causes stimulated transitions
from excited to ground states and vice versa. The
corresponding rate is proportional to the
spectral density of the radiation taken at the atomic transition frequency.
Apart from some dimensional constants
 ${\widetilde D}_{\mu\nu}^{-+}$
is precisely the spectral energy density of the incoherent field
at the position $\vec r_0$ and at the 
transition frequency $\omega$ of the probe atom. 
\begin{equation}
\gamma_{\mu\nu}(\omega,t)=\frac{\wp_\mu\wp_\nu}{\hbar^2}\int_{-\infty}^\infty\!\!\! 
d\tau \Bigl\langle [ E_\nu^+(\vec r_0,t+\tau),E_\mu^-(\vec r_0,t)]\Bigr\rangle
\, e^{i\omega\tau}\label{Einstein_A}
\end{equation}
is the spontaneous contribution to the ``down rate''
in the atomic medium. 
(Note that the commutator contains the field operators {\it interacting
with the environment atoms.})
Since we are not interested here in the
effects of the medium to the spontaneous decay, we replace $\gamma_{\mu\nu}$
by the free-space value $\gamma_{\mu\nu}^0$.
We will show in Appendix A, that Eq.(\ref{Einstein_A}) 
indeed leads to the well-known Wigner-Weisskopf result for radiative decay 
in free space, if we replace $E$ by the free field.
Light-shifts induced by the
incoherent component of the radiation field inside the medium are described by 
the hermitian matrix
\begin{eqnarray}
\lefteqn{H_{\mu\nu}(\omega,t)\;=}  \nonumber \\
&& \frac{i}{\hbar}\frac{\wp_\mu\wp_\nu}{2} 
\int_0^\infty\!\!\!
d\tau \biggl[\langle\langle E_\mu^-(\vec r_0,t) E^+_\nu(\vec r_0,t-\tau)
\rangle\rangle\, e^{-i\omega\tau} - \nonumber \\
&& \qquad \langle\langle E_\mu^-(\vec r_0,t) E^+_\nu(\vec r_0,t+\tau)
\rangle\rangle\, e^{+i\omega\tau}\biggr].\label{H_1}
\end{eqnarray}
Eq.(\ref{H_1}) can also be expressed in terms of $D^{-+}$:
\begin{eqnarray}
H_{\mu\nu}(\omega,t)
&=& \frac{\wp_\mu\wp_\nu}{2\pi\hbar}
\, {\rm P}\!\!\int_{-\infty}^\infty\!\!\!d\omega^\prime
\enspace\frac{{\widetilde D}_{\mu\nu}^{-+}(\vec r_0,\omega^\prime;t)}
{\omega-\omega^\prime} \nonumber \\
&=& \qquad \frac{\hbar}{2\pi}\, {\rm P}\!\!\int_{-\infty}^\infty 
\!\!\! d\omega^\prime
\enspace\frac{\Gamma_{\mu\nu}(\omega^\prime,t)}{\omega-\omega^\prime},
\end{eqnarray}
where { P}~denotes the principle part of the integral.
In systems with inhomogeneous broadening the collective light-shifts
are often negligible as they are usually small compared to the
inhomogeneous width.     
\begin{eqnarray}
\lefteqn{h_{\mu\nu}(\omega, t)=
\frac{i}{\hbar}\frac{\wp_\mu\wp_\nu}{2} \int_0^\infty\!\!\!
d\tau} \\
&&\qquad \biggl[
\langle [E_\mu^-(\vec r_0,t),E_\nu^+(\vec r_0,t-\tau)]\rangle\, 
e^{-i\omega\tau} - \nonumber \\
&& \qquad \langle [E_\mu^-(\vec r_0,t),E_\nu^+(\vec r_0,t+\tau)]\rangle\, 
e^{i\omega\tau}\biggr]
\end{eqnarray}
is the corresponding spontaneous contribution. Within the approximations made,
$h_{\mu\nu}$ reflects the Lamb-shift of excited states altered by the 
presence of 
the medium. Here we are not interested in the Lamb shift 
and therefore consider it included in the free Hamiltonian $H_0$.

The effective time-evolution operator (\ref{Seff_RWA2})
directly leads to the following master equation for the single-atom
density operator:
\begin{eqnarray}
\dot\rho &=& -\frac{i}{\hbar}\Bigl[H_0,\rho\Bigr]+
i\frac{\wp_\mu}{\hbar}\biggl[\sigma_\mu {\cal E}_{L\mu}^- +
\sigma_\mu^\dagger {\cal E}_{L\mu}^+,\rho\biggr]
\nonumber\\
&& + \frac{i}{\hbar} H_{\mu\nu}\biggl[\sigma_\mu\sigma_\nu^\dagger -
\sigma_\nu^\dagger\sigma_\mu,\rho\biggr]\label{DME}\\
&& - \frac{\Gamma_{\mu\nu}}{2}
\biggl\{ \sigma_\mu\sigma_\nu^\dagger \rho + \rho \sigma_\mu\sigma_\nu^\dagger
-2 \sigma_\nu^\dagger \rho\sigma_\mu\biggr\}\nonumber\\
&& - \biggl(\frac{\Gamma_{\mu\nu}}{2}+\frac{\gamma^0_{\mu\nu}}{2}\biggr)
\biggl\{\sigma_\nu^\dagger \sigma_\mu \rho + \rho \sigma_\nu^\dagger \sigma_\mu
-2 \sigma_\mu\rho\sigma_\nu^\dagger  \biggr\}.\nonumber
\end{eqnarray}
This is the first main result of the present paper. We note that this
equation is nonlinear and nonlocal, since
the light-shift and decay matrices depend via the field correlations
on the surrounding atoms. The equation does however have the Lindblad form
\cite{Lindblad}
and thus preserves positivity and the total probability. 
In order to obtain a closed set of equations, we calculate in
the following
sections the yet unknown quantities ${\cal E}_L$, 
$\Gamma_{\mu\nu}$, and $H_{\mu\nu}$ in terms of single-atom density matrix
elements.


\section{the average local field and the Lorentz-Lorenz relation}


We recognize from Eq.(\ref{DME}) that the probe atom is coupled to a classical
(c-number) field of amplitude
\begin{equation}
\vec{\cal E}_L(\vec r,t) =\vec{\cal E}(\vec r,t) +\langle\vec
 E(\vec r,t)\rangle
\label{local}.
\end{equation}
The first term is the external coherent field
( = field in the absence of the medium), and the second term
is the mean coherent amplitude of the field scattered by all other atoms.
Note, that the contribution of the probe atom itself is not included.
On the other hand, the macroscopic field ${\cal E}_M$, which enters Maxwell's 
equations is the {\it total}
field inside the medium (averaged over a spatial region large compared to
the characteristic atomic distance, but smaller than $\lambda^3$).
Thus the {\it local} field, given in (\ref{local}) differs from the
macroscopic Maxwell field essentially by the scattering 
contribution of the probe atom itself. 
In a continuum approximation we find
\begin{eqnarray}
\lefteqn{{\cal E}_{L\alpha}(\vec r,t) \;=\; {\cal E}_{M\alpha}(\vec
r,t) -} \nonumber \\
&& \frac{i}{\hbar} \varrho \int_{K_\epsilon}\!\! d^3{\vec r\,}^\prime 
\int_{-\infty}^\infty\!\!\! dt^\prime \, D_{0\, \alpha\beta}^{\rm ret}
(\vec r,t;{\vec r\,}^\prime,t^\prime)\, \langle p_{H\beta}(t^\prime)\rangle,
\label{LL02}
\end{eqnarray}
where $D_0^{\rm ret}$ is the free-field retarded propagator and 
$\langle  p_{H}\rangle$ is the expectation value of the probe-dipole operator
(in the Heisenberg picture). $\varrho$ is the atomic density and $K_\epsilon$
denotes integration over a small sphere of radius $\epsilon$.
The retarded propagator of the electric field is given by \cite{Pauli}
\begin{eqnarray}
&&D_{0\, \alpha\beta}^{\rm ret}(1,2)=\label{GFret}\\
&&\quad\quad\frac{i\hbar}{4\pi\epsilon_0 c}\Theta(\tau)
\left[\delta_{\alpha\beta} \frac{\partial^2}{\partial \tau^2}-c^2\frac{
\partial^2}{\partial x_2^\alpha\partial x_2^\beta}\right]
\frac{\delta(r-c\tau)} {r},\nonumber
\end{eqnarray}
where $\tau=t_1-t_2$, $r=\vert \vec r_1 -\vec r_2\vert$ and $\Theta$ is
the Heaviside step function.

When substituting $D_0^{\rm ret}$ from Eq.(\ref{GFret}) into Eq.(\ref{LL02})
we note that in the limit $\epsilon\to 0$ only the term which results from 
the second spatial derivative of $1/r$ survives. Using
\begin{equation}
\frac{\partial^2}{\partial x_2^\alpha\partial x_2^\beta}\enspace\frac{1}{r}=
-\frac{4\pi}{3}\enspace \delta^{(3)}(\vec r_1-\vec r_2)
\enspace\delta_{\alpha\beta},
\end{equation}
we find
\begin{eqnarray}
{\vec{\cal E}}_L(\vec r,t)&=&{\vec{\cal E}}_M(\vec r,t) 
+\frac{1}{3 \epsilon_0}\varrho
 \Bigl\langle \vec p_H (t)\Bigr\rangle  \nonumber \\
&=&  {\vec{\cal E}}_M(\vec r,t) 
+\frac{1}{3 \epsilon_0}{\vec{\cal P}}
\label{Lorentz}
\end{eqnarray}
which is identical to the classical Lorentz-Lorenz relation \cite{LL,LL2}
when we identify $\vec{\cal P}=\varrho\langle\vec p_H\rangle$. It should be
mentioned
that the Lorentz-Lorenz relation holds for the mean amplitude of the field 
and not for the field operators itself as claimed in \cite{Burnett}. 

Making use of (\ref{Lorentz}) we can define an effective semiclassical
interaction operator
\begin{equation}
V_L=
-\sum_j {\vec p}_\mu^{\, j}(t) {\vec {\cal E}}_{L\mu}(\vec r_j,t) \quad .
\label{VL}
\end{equation}


\section{quantum corrections}


We now discuss the light-shift and decay matrices 
in the generalized density-matrix equation
(\ref{DME}) in more detail. Both depend on the field cumulants
or Greensfunctions
\begin{equation}
D_{\mu\nu}^{-+}(\vec r,t;\vec r,t^\prime)=
\langle\langle E_\mu^-(\vec r,t) E_\nu^+(\vec r,t^\prime)\rangle\rangle.
\end{equation}
(Note that the superscript ``$-+$'' indicates
 that the first time argument is on the
lower and the second time argument on the upper branch of the Keldysh contour
and has nothing to do with the frequency components of the field.)
The aim of the present section is to calculate $D^{-+}$
in terms of atomic variables.
For this we apply non-equilibrium Greensfunction techniques \cite{Fetter}.


\subsubsection{Dyson-equation for $D(\check 1,\check 2)$}


We define the exact and the (known) 
free Greensfunctions (GF) on the Keldysh contour as
\begin{eqnarray}
D_{\mu\nu}(\check 1,\check 2)&=& \langle\langle T_C 
E_\mu^-(\vec r_1,\check t_1)E_\nu^+(\vec r_2,\check t_2)\rangle\rangle,\\
D_{0\,\mu\nu}(\check 1,\check 2)&=& \langle\langle T_C 
E_{0\mu}^-(\vec r_1,\check t_1)E_{0\nu}^+(\vec r_2,\check t_2)\rangle\rangle,
\end{eqnarray}
where $E_0$ denotes the free field, i.e.~without coupling to the medium.
The contour-Greensfunction
$D(\check 1,\check 2)$ contains four real-time GFs:
$D^{++}(1,2)$, $D^{-+}(1,2)$, $D^{+-}(1,2)$, and $D^{--}(1,2)$,
where the superscripts ``$\pm$'' specify contour branches. 
The first and the last are the time- and anti-time ordered propagators
and the retarded and advanced propagators are given by the combinations
\cite{Fetter}
\begin{eqnarray}
D^{\rm ret}(1,2) &=& D^{++}(1,2) - D^{+-}(1,2)\\
                 &=& D^{-+}(1,2) - D^{--}(1,2),\nonumber\\
D^{\rm adv}(1,2) &=& D^{++}(1,2) - D^{-+}(1,2)\\
                 &=& D^{+-}(1,2) - D^{--}(1,2).\nonumber
\end{eqnarray}
Within the RWA and in the absence of thermal photons, we have
\begin{eqnarray}
D_{0\alpha\beta}^{++}(1,2) &\approx& D_{0\alpha\beta}^{\rm adv}(1,2)_,
\label{D0++RWA}\\
D_{0\alpha\beta}^{-+}(1,2) &\approx& 0,
\label{D0-+RWA}\\
D_{0\alpha\beta}^{+-}(1,2) &\approx& D_{0\alpha\beta}^{\rm adv}(1,2)
                              -D_{0\alpha\beta}^{\rm ret}(1,2),
\label{D0+-RWA}\\
D_{0\alpha\beta}^{--}(1,2) &\approx&-D_{0\alpha\beta}^{\rm ret}(1,2).
\label{D0--RWA}
\end{eqnarray}

A formal solution to the atom-field interaction can be given in terms
of a Dyson-integral equation \cite{Fetter},
by introducing
a formal polarization function $\Pi_{\alpha\beta}(\check 1,\check 2)$
\begin{eqnarray}
\lefteqn{D_{\mu\nu}(\check 1,\check 2) \;=\; D_{0\, \mu\nu} (\check
1,\check 2) -} \nonumber \\
&& \int\!\!\!\int_C\! d\check 1^\prime\,d\check 2^{\, \prime}\,
D_{0\, \mu\alpha}(\check 1,\check 1^\prime)\, 
\Pi_{\alpha\beta}(\check 1^\prime,\check 2^{\, \prime})\, D_{\beta\nu}
(\check 2^{\, \prime},\check 2).\label{Dyson}
\end{eqnarray}
Here $\int_C d\check 1 $ denotes integration over the 
Schwinger-Keldysh contour as well as spatial integration over the
medium. The Dyson-equation (\ref{Dyson}) represents 
nothing else than a formal summation
of the perturbation series where the polarization function 
is determined by the medium response. We now have to find a 
good approximation for $\Pi(1,2)$.


\subsubsection{self-consistent Hartree approximation}


One easily verifies that in lowest order in the atom-field coupling,
the polarization function is given by 
a correlation function of dipole operators 
of non-interacting atoms
\begin{eqnarray}
\lefteqn{\Pi^{(0)}_{\alpha\beta} (\check 1,\check 2) \;=} \label{Hartree} \\
&& \frac{\wp_\alpha\wp_\beta}{\hbar^2}
\sum_j \bigl\langle\bigl\langle 
T_C \sigma^\dagger_{j\alpha}(\check t_1) \sigma_{j\beta}
(\check t_2)\bigr\rangle\bigr\rangle_{\rm free}\, \delta(\vec r_1-\vec r_j)\,
\delta(\vec r_2-\vec r_j). \nonumber
\end{eqnarray}
This corresponds to a Hartree approximation in many-body theory. 
This approximation
is physically justified, when the nonlinear light-shift and
decay terms do not affect the atomic dynamics, that is if the 
probability that a specific atom reabsorbs or scatters a spontaneous
photon is small. Such a situation is realized, for example, 
in the classical case of radiation trapping where a small number of 
photons (much smaller than
necessary to saturate the medium) is trapped in a dense absorbing medium
\cite{Holstein}. We are here also interested, however, in 
situations, where incoherent photons  significantly alter
the atomic dynamics. A consistent approximation, which accounts also
for these cases is the self-consistent Hartree approximation, where the
cumulants of {\it free} dipole operators in  (\ref{Hartree})
are replaced by cumulants of {\it interacting} dipole operators.
\begin{eqnarray}
\lefteqn{\Pi_{\alpha\beta} (\check 1,\check 2) \;=}\nonumber \\
&& \frac{\wp_\alpha\wp_\beta}{\hbar^2}
\sum_j \langle\langle T_C \sigma^\dagger_{j\alpha}(\check t_1) 
\sigma_{j\beta}
(\check t_2)\rangle\rangle\, \delta(\vec r_1-\vec r_j)\,
\delta(\vec r_2-\vec r_j).\label{Hartree_sc}
\end{eqnarray}

As shown in Appendix B the Dyson-equation for the contour GF
can be rewritten in the RWA in terms of the real-time GFs as follows:  
\begin{equation}
D^{-+}_{\alpha\beta}(1,2) = -\int\!\!\!\int d3\, d4\, 
D^{\rm ret}_{\alpha\mu}(1,3)\, 
\Pi^{\, \rm s}_{\mu\nu}(3,4)\,
D^{\rm adv}_{\nu\beta}(4,2),\label{GF_eq1}
\end{equation}
where $D^{\rm ret}_{\mu\nu}(1,2)\left(=D_{\nu\mu}^{\rm adv}(2,1)\right)$
obeys the Dyson equation
\begin{eqnarray}
\lefteqn{D^{\rm ret}_{\alpha\beta}(1,2) \;=\; D_{0\,\alpha\beta}^{\rm
ret}(1,2) -} \nonumber \\
&& \int\!\!\!\int d3\,d4\, 
 D_{0\,\alpha\mu}^{\rm ret}(1,3)\, \Pi^{\rm ret}_{\mu\nu}(3,4)\, 
D^{\rm ret}_{\nu\beta}(4,2).
\label{GF_eq2}
\end{eqnarray}
Here the time integration goes from $-\infty$ to $\infty$ and
we have introduced the atomic source correlation
\begin{eqnarray}
\lefteqn{\Pi^{\, \rm s}_{\mu\nu}(\vec r_1,t_1;\vec r_2,t_2) \;=}
\nonumber \\
&& \frac{\wp_\mu\wp_\nu}{\hbar^2}
\sum_j \bigl\langle\bigl\langle \sigma^\dagger_{j\mu}(t_1) \sigma_{j\nu}
(t_2)\bigr\rangle\bigr\rangle\, \delta(\vec r_1-\vec r_j)\,
\delta(\vec r_2-\vec r_j)\label{source}
\end{eqnarray}
as well as the atomic response function 
\begin{eqnarray}
\lefteqn{\Pi^{\rm ret}_{\mu\nu}(\vec r_1,t_1;\vec r_2,t_2) \;=\;
\frac{\wp_\mu\wp_\nu}{\hbar^2}
\Theta(t_1-t_2)} \nonumber \\
&& \sum_j 
\bigl\langle \bigl[\sigma_{j\mu}^\dagger(t_1), \sigma_{j\nu}
(t_2)\bigr]\bigr\rangle\, \delta(\vec r_1-\vec r_j)\,
\delta(\vec r_2-\vec r_j).\label{response}
\end{eqnarray}
The  names reflect the physical meaning of the terms.
The Fourier-transform of $\Pi^{\, \rm s}$ is proportional to
the spontaneous emission spectrum of the atoms and that of $\Pi^{\rm ret}$
gives the susceptibility of the medium. 

\begin{figure}
\epsfxsize=4cm
\centerline{\leavevmode\epsffile{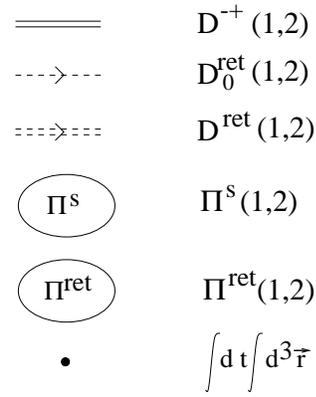}}
\caption{Feynman diagrams, definitions}
\end{figure}

\begin{figure}
\epsfxsize=4cm
\centerline{\leavevmode\epsffile{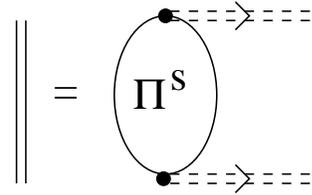}}
\caption{Graphical representation of Eq.(\ref{GF_eq1}). The incoherent 
intensity at the position of the probe atom is the sum of all spontaneous
contribution propagated through the medium.}
\end{figure}

\begin{figure}
\epsfxsize=6cm
\centerline{\leavevmode\epsffile{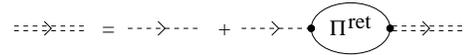}}
\caption{Graphical representation of Dyson equation (\ref{GF_eq2}) 
for retarded
GF inside the medium. Iteration generates all-order scattering
contributions.}
\end{figure}

Eqs.(\ref{GF_eq1}) and (\ref{GF_eq2}) can be given an
instructive graphical representation shown in Figs.~2-4. 
Eq.(\ref{GF_eq1}) (illustrated in Fig.~3) 
says that the incoherent radiation intensity
is obtained by summing the spontaneous-emission contributions from 
all atoms propagated through the medium. 
The iteration of the Dyson equation (\ref{GF_eq2})
(shown in Fig.~4) describes multiple scattering of  spontaneous photons
by atoms during the propagation from a source atom to the probe atom.


\subsubsection{Explicit expressions for the collective decay rate and 
light-shift}


We now approximately  solve the Dyson-equation (\ref{GF_eq2}) for the
retarded propagator in the medium. We first introduce a continuum 
approximation.
\begin{eqnarray}
\lefteqn{ \Pi_{\mu\nu}^{\rm ret}({\vec r\,}_1,t_1;{\vec r\,}_2t_2)\;=}
\nonumber \\
&&\int\!\! d^3\vec r\enspace P^{\rm ret}_{\mu\nu}(\vec r,t_1,t_2)
\, \delta(\vec r_1-\vec r)\, \delta(\vec r_2-\vec r),
\end{eqnarray}
\begin{eqnarray}
\lefteqn{P^{\rm ret}_{\mu\nu}(\vec r_j,t_1,t_2) =} \nonumber \\
&& \frac{\wp_\mu\wp_\nu}{\hbar^2}N  
\Theta(t_1-t_2)\, \overline{\, \langle[\sigma_{j\mu}^\dagger(t_1),\sigma_{j\nu}(t_2)]
\rangle\,},
\end{eqnarray}
 where $N$ is the atom density and the overline denotes averaging over
some inhomogeneous distribution.
Similarly
\begin{eqnarray}
\lefteqn{ \Pi_{\mu\nu}^{\rm\, s}({\vec r\,}_1,t_1;
{\vec r\,}_2,t_2)\;=} \nonumber \\
&&\int\!\! d^3\vec r\enspace P^{\rm \, s}_{\mu\nu}(\vec r,t_1,t_2)
\, \delta(\vec r_1-\vec r)\, \delta(\vec r_2-\vec r),\\
\lefteqn{P^{\rm\, s}_{\mu\nu}(\vec r_j,t_1,t_2) =
\frac{\wp_\mu\wp_\nu}{\hbar^2}N 
\, \overline{\,\langle\langle\sigma_{j\mu}^\dagger(t_1)\sigma_{j\nu}(t_2)
\rangle\rangle\,}},
\end{eqnarray}
Thus Eq.(\ref{GF_eq2}) reads
\begin{eqnarray}
\lefteqn{D_{\alpha\beta}^{\rm ret}(\vec r_1,t_1;\vec r_2,t_2)
\;=\; D_{0\,\alpha\beta}^{\rm ret}(\vec r_1,t_1;\vec
r_2,t_2)\label{Dyson_expl} -}\\ 
&&\quad
\int_{-\infty}^\infty\!\!\! dt_1^\prime\int_{-\infty}^\infty\!\!\! 
dt_2^\prime\int_V
\! d^3{\vec r\,}_1^\prime\,\, 
D_{0\,\alpha\mu}^{\rm ret}(\vec r_1,t_1;{\vec
r\,}_1^\prime,t_1^\prime) \nonumber \\
&&\qquad\qquad P_{\mu\nu}^{\rm ret}({\vec r\,}_1^\prime;t_1^\prime,t_2^\prime)
\,\, D_{\nu\beta}^{\rm ret}({\vec r\,}_1^\prime,t_2^\prime;\vec r_2,t_2).
\nonumber
\end{eqnarray}
To solve this integral equation, we now make the following
approximations. We first extend the spatial integration 
to infinity, which basically means that we are
solving for the retarded propagator in an infinitely extended medium. 
Secondly, we replace ${\vec r\,}_1^\prime$ in the atomic response function
by $\vec r_2$, i.e. we evaluate the response at the position of the source.
We furthermore consider quasi-stationary conditions, i.e.~assume that
$P^{\rm ret}(t_1^\prime,t_2^\prime)$ depends only on the time 
difference $\tau=t_1^\prime-t_2^\prime$.
We only keep an overall slow (parametric) time dependence.
This means that we consider propagation times short compared to the
characteristic time of the atomic evolution, which is consistent with the
earlier Markov approximation. 
With these simplifications we can turn the
integral equation (\ref{Dyson_expl}) into an algebraic one by 
Fourier-transformation with respect to $\vec x\equiv \vec r_1-\vec r_2$ and
$\tau\equiv t_1-t_2$. At this point a word of caution is needed: 
As will be discussed
in Appendix C, the retarded GF in an {\it amplifying} medium is not 
Fourier-transformable, since it grows exponentially with $r=|\vec x|$.
We therefore should view  the transformations as finite-time and
finite-space Fourier-transforms, and hence the algebraic equation as an 
approximation. 

Using the definition
\begin{equation}
{\widetilde{\widetilde {\rm F}}}(\vec q,\omega)
=\int_{V_\infty}\!\!\! d^3\vec x\int_{-\infty}^\infty\!\!\!
 d\tau\, {\rm F}(\vec x,\tau) 
\, e^{-i\omega\tau}\, e^{i\vec q\cdot\vec x},
\end{equation}
the solution of Eq.(\ref{Dyson_expl}) reads
\begin{equation}
\widetilde{\widetilde{\bf D}}^{\rm ret}(\vec q,\omega;t)
=\left[{\bf 1} + \widetilde{\widetilde{\bf D}}^{\rm ret}_0(\vec q,\omega)
\cdot {\widetilde{\bf P}}^{\rm ret}(\omega;t)\right]^{-1}
\cdot \widetilde{\widetilde{\bf D}}^{\rm ret}_0(\vec q,\omega).
\label{Dyson_matrix}
\end{equation}
Here ${\bf D}^{\rm ret}$ and ${\bf \Pi}^{\rm ret}$ denote $3\times3$ 
matrices in coordinate space and ${\bf 1}$ is the unity matrix.

For simplicity we now disregard polarization, i.e.~we replace the $3\times3$
matrices by simple functions. We note however, that a generalization
is  straight forward. As shown in detail in Appendix C, we 
eventually arrive at
\begin{eqnarray}
{\widetilde{D}}^{\rm ret}(\vec x,\omega;t)
= -\frac{i\hbar\omega^2}{6 \pi\epsilon_0 c^2} 
\frac{e^{q_0^{\prime\prime} r}}{r}
\, e^{-i q_0^\prime r} \label{Dyson_sol}
\end{eqnarray}
where $\lambda$ is the wavelength of the
transition under consideration in the rest frame,
and $r=|\vec x|=|\vec r_1-\vec r_2|$.
\begin{equation}
q_0=q_0^\prime(\vec r,\omega,t)+ iq_0^{\prime\prime}(\vec r,\omega,t) 
= \frac{\omega}{c}
\left[1+ \frac{i\hbar }{3\epsilon_0} {\widetilde P}^{\rm ret}(\vec r,\omega;t)
\right].\label{q_0}
\end{equation}
$q_0^{\prime\prime}$ is the inverse absorption/amplification length 
in the medium and $q_0^{\prime}$ characterizes the corresponding
phase shift. We here have assumed that $\bigl|{\rm Im} 
[{\widetilde P}^{\rm ret}]\bigr|\, \hbar/3\epsilon_0 < 1$. 

With Eq.(\ref{Dyson_sol}) we can now express ${\widetilde D}^{-+}
(\vec r_0,\omega;t)$
in terms of atomic variables
\begin{equation}
{\widetilde D}^{-+}
(\vec r_0,\omega,t)=
\frac{ \hbar^2\omega^4 }{(6\pi)^2\epsilon_0^2 c^4}
\int_V\! d^3\vec r \enspace \frac{e^{2 q_0^{\prime\prime}
(\vec r,\omega;t) r}}{r^2}
\, {\widetilde P}^{\rm \, s} (\vec r,\omega;t).
\label{D-+sol}
\end{equation}
Here $r=|\vec r-\vec r_0|$ is the distance between source and probe atom.
With Eq.(\ref{D-+sol}) we finally find for the collective decay rate
and light-shift 
\begin{eqnarray}
\Gamma(\omega,t) &=& 
\frac{\wp^2 \omega^4 }{(6 \pi)^2 \epsilon_0^2 c^4}
\int_V\! d^3\vec r \, \frac{e^{2 q_0^{\prime\prime}(\vec r,\omega;t) r}}{r^2}
\, {\widetilde P}^{\rm \, s} (\vec r,\omega;t),
\label{G_sol}\\
H(\omega,t) &=& 
\frac{\hbar\wp^2 \omega^4 }{(6 \pi)^2 \epsilon_0^2 c^4}
\int_V\! d^3\vec r \nonumber \\
&& \qquad\qquad {\rm P}\!\!\int_{-\infty}^\infty 
\frac{d\omega^\prime}{2\pi}
 \, \frac{e^{2 q_0^{\prime\prime}(\vec r, \omega^\prime;t) r}}{r^2}
\,\, \frac{{\widetilde P}^{\rm \, s} (\vec r,\omega;t)}
{\omega-\omega^\prime}.\label{H_sol}
\end{eqnarray}
Eqs.(\ref{G_sol}) and (\ref{H_sol}) are the second major result of the
present paper. In applying these results to a specific problem, we still
have to calculate the source-correlation $P^{\rm\, s}$ in terms of density
matrix elements. This then yields a closed nonlinear and nonlocal density
matrix equation. We will illustrate this for some examples in the following
section.


\section{examples}


\subsection{Inhomogeneously broadened two-level system}


We here consider an inhomogeneously broadened  dense ensemble
of randomly polarized two-level atoms in a cylindrical geometry 
as shown in Fig.~5.
For this system
the time-evolution of the dipole operator $\sigma=|b\rangle\langle a|$ 
is determined by the simple Heisenberg-Langevin equation
\begin{equation}
\dot\sigma_j = -(i\omega_{ab}^j +\Gamma_{ab})\sigma_j 
+ {\rm noise},\label{sigma}
\end{equation}
where the noise term denotes a white noise source, which is however of no
interest here. $\omega_{ab}^j=\omega_{ab}^0+\Delta_j$ is the atomic transition 
frequency in the laboratory frame. We here take into account Doppler-broadening
which leads to a shift $\Delta_j$ of the lab-frame transition frequency 
from the rest-frame frequency $\omega_{ab}^0$. The collective
light-shift is  small compared to the average Doppler-shift
and therefore neglected. 
The coherence decay rate $\Gamma_{ab}$ consists of two 
contributions, one resulting from free-space spontaneous decay $\gamma$
and the other from the collective decay $\Gamma$, $\Gamma_{ab}=\Gamma
+\gamma/2$.
 Eq.(\ref{sigma}) can easily be solved by Laplace-transformation
($\tilde x(s,t) :=\int_0^\infty d\tau\, e^{-s\tau}\, x(t+\tau)$), which 
yields
\begin{eqnarray}
\langle\langle\tilde\sigma^\dagger_j(s;t)\sigma_j(t)\rangle\rangle &=&
\frac{\rho_{aa}^j(t)}{s-i\omega_{ab}^j+\Gamma_{ab}},\\
\langle\langle\tilde\sigma_j(s;t)\sigma_j^\dagger(t)\rangle\rangle &=&
\frac{\rho_{bb}^j(t)}{s+i\omega_{ab}^j+\Gamma_{ab}}.
\end{eqnarray}
From this we immediately obtain
\begin{eqnarray}
{\widetilde P}^{\rm ret}(\vec r_j,\omega,t)&=& \frac{\wp^2}{\hbar^2}
N\overline{\, \frac{\rho_{aa}^j(t) -\rho_{bb}^j(t)}{\Gamma_{ab}
+i(\omega-\omega_{ab}^j)}\, },\label{Pi_ret_TLA}\\
&&\nonumber\\
{\widetilde P}^{\rm\, s}(\vec r_j,\omega,t)&=& \frac{2\wp^2}{\hbar^2}
N\overline{\, \frac{\rho_{aa}^j(t)\Gamma_{ab}}{(\Gamma_{ab})^2
+(\omega-\omega_{ab}^j)^2}\, },
\end{eqnarray}
where $\wp$ is the dipole moment of the transition and the overbar
 denotes averaging over the velocity
distribution of the atoms, which is given by the Gaussian distribution
\begin{equation}
W(\Delta_j) = \frac{1}{\sqrt{2\pi}\Delta_D} \exp\left\{ -\frac{\Delta_j^2}
{2 \Delta_D^2}\right\}.\label{W}
\end{equation}
\begin{figure}
\epsfxsize=6cm
\centerline{\leavevmode\epsffile{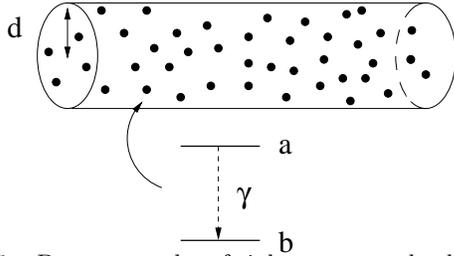}}
\caption{Dense sample of inhomogeneously broadened 
two-level atoms in cylindrical geometry.}
\end{figure}

Since the lab-frame atomic transition frequency depends on the velocity,
the collective decay rate, which is proportional to the incoherent 
radiation spectrum
at this frequency,  will be velocity dependent as well. Thus we have
in general a set of nonlinear coupled equations corresponding to
different velocity classes. If there are  fast velocity-changing
collision, the population dynamics of all velocity classes will however be 
approximately the same. 
In this case we may set
\begin{equation}
\rho_{\mu\mu}^j(t) = \overline{\, \rho_{\mu\mu}^j(t)\, } =:
\rho_{\mu\mu}(\vec r,t),
\label{fast_coll_appr}
\end{equation}
where $\mu\in\{a,b\}$ and $\vec r$ denotes the position of the atoms 
considered. 
Note, however, that this approximation does not hold
if the inhomogeneous broadening mechanism is not due to Doppler as for 
example in solids. In that case one has to consider the full set of
equations. Using Eq.(\ref{fast_coll_appr}), we find in the limit of
large Doppler-broadening
$\Delta_D\gg\Gamma_{ab}$
\begin{eqnarray}
{\widetilde P}^{\rm ret}(\vec r,\omega,t)&=& \frac{\wp^2 N}{\hbar^2}
\sqrt{2\pi}\, \frac{\rho_{aa}(\vec r,t)-\rho_{bb}(\vec r,t)}
{\Delta_D} \nonumber \\
&& e^{-\Delta^2/2\Delta_D^2}\, \biggl[ 1 - i \sqrt{\frac{2}{\pi}}
\int_0^{\Delta/\Delta_D}\!\!\! dy\, e^{-y^2/2}\biggr],
\end{eqnarray}
where $\Delta=\omega-\omega_{ab}^0$ is the detuning from the atomic resonance
at rest. For the collective decay rate only the real part of 
${\widetilde P}^{\rm ret}$
is important which enters the absorption coefficient according to 
 Eq.(\ref{q_0})
\begin{eqnarray}
\lefteqn{q_0^{\prime\prime}(\vec r,\omega,t) \;=}\label{q_0_TLA} \\
&&  \frac{\wp^2 N}{3\hbar\epsilon_0}
\frac{\omega}{c} \sqrt{2\pi}\, \frac{\rho_{aa}(\vec r,t)
-\rho_{bb}(\vec r,t)}{\Delta_D}
\, e^{-\Delta^2/2\Delta_D^2}.\nonumber 
\end{eqnarray}
Similarly we have for the source term in the strong Doppler-limit
\begin{equation}
{\widetilde P}^{\rm\, s}(\vec r,\omega,t) = 
\frac{2\wp^2 N}{\hbar^2}
\sqrt{2\pi}\, \frac{\rho_{aa}(\vec r,t)}{\Delta_D}
\, e^{-\Delta^2/2\Delta_D^2}.\label{Pi_s_TLA}
\end{equation}

Combining (\ref{q_0_TLA}) and (\ref{Pi_s_TLA}) 
and applying the relation between the free-space radiative 
decay rate $\gamma$ and the dipole moment $\wp$: 
$\wp^2 = 3\pi\hbar\epsilon_0 c^3 \gamma/\omega^3$
\cite{Louisell} (cf.~also Appendix A)
yields the collective decay rate  (\ref{G_sol}) 
for a probe atom with (lab-frame) transition frequency 
$\omega$ at position $\vec r_0$
\begin{eqnarray}
\lefteqn{\Gamma(\omega,t)\;=} 
\label{G_TLA_omega}\\
&& \gamma \, \int_V\! d^3\vec r\, \, 
2 q_0^{\prime\prime}(\vec r,\omega,t) 
\, \frac{e^{2 q_0^{\prime\prime}(\vec r,\omega,t)r}
}{4\pi r^2}\, 
\frac{\rho_{aa}(\vec r,t)}{\rho_{aa}(\vec r,t)-\rho_{bb}(\vec r,t)},
\nonumber 
\end{eqnarray}
with $r=|\vec r-\vec r_0|$.
To obtain the effective decay/pump rate we have to 
averaged over the velocity distribution
\begin{eqnarray}
\lefteqn{\Gamma(t)\;=\;\overline{\, \Gamma(\omega,t)\, } \;=}
\label{G_TLA_av} \\
&& \int_{-\infty}^\infty \!\!\!d\omega\, \frac{1}{\sqrt{2\pi}\Delta_D}
e^{-(\omega-\omega_{ab}^0)^2/2\Delta_D^2} \,
\Gamma(\omega,t). \nonumber 
\end{eqnarray}

We now discuss two limiting cases. In the first case we assume a small
excitation in the medium. This corresponds to the classical situation of
radiation trapping in an inhomogeneously broadened two-level
medium. We will show that in this case Eq.(\ref{G_TLA_av})
leads to the integral equation of Holstein \cite{Holstein}. 
In the second case we will disregard the spatial dependence but keep the 
nonlinearities, and consider
the temporal evolution from an initially  excited ensemble.


\subsubsection{linear limit and Holstein equations of radiation trapping}


For small excitation, the retarded light propagation can be regarded as
propagation in a medium with all population in the lower state, i.e.
$\rho_{bb}=1$ and $\rho_{aa}=0$. Thus
\begin{equation}
q_0^{\prime\prime}(\vec r,\omega,t) =q_0^{\prime\prime}(\omega,t)
= -\frac{\wp^2 N}{3\hbar\epsilon_0}\frac{\omega}{c} 
\frac{\sqrt{2\pi}}{\Delta_D} \, e^{-\Delta^2/2 \Delta_D^2}
\end{equation}
and we can approximate the denominator in (\ref{G_TLA_omega})
by $-1$. This results in
\begin{equation}
\Gamma(t)\approx
\gamma \,  \int_V\!\! d^3\vec r\enspace G(\vec r_0,\vec r)
\, \rho_{aa}(\vec r,t)\label{G_TLA_lin}
\end{equation}
where
\begin{eqnarray}
\lefteqn{G(\vec r_0,\vec r) \;=} \\
&& - \overline{\, 2 q_0^{\prime\prime}(\omega,t)
 \, \frac{e^{2 q_0^{\prime\prime}(\omega,t)r}}
{4\pi r^2}\,}= \nonumber \\
&&
-\int_{-\infty}^\infty \!\!\!d\omega\, \frac{1}{\sqrt{2\pi}\Delta_D}
e^{-(\omega-\omega_{ab}^0)^2/2\Delta_D^2}\,   2 q_0^{\prime\prime}(\omega,t)
 \, \frac{e^{2 q_0^{\prime\prime}(\omega,t)r}}
{4\pi r^2} = \nonumber \\
&& \frac{1}{\sqrt{\pi}} \int_{-\infty}^\infty \!\! dx\, e^{-x^2}
\left(-\frac{1}{4\pi r^2}\right) \frac{\partial}{\partial r}
\exp\Bigl[-K_0 e^{-x^2} r\Bigr]. \nonumber 
\end{eqnarray}
Here $K_0=N\lambda^2\,g$ and 
$g=\gamma/\sqrt{2\pi}\Delta_D$ characterizes the ratio of
the homogeneous to the inhomogeneous width.

The dynamical evolution of the ensemble is 
described by the Bloch equation
\begin{equation}
\dot\rho_{aa}(\vec r_0,t) = \Gamma(t) - 
\Bigl[\gamma + 2 \Gamma(t)\Bigr]\, \rho_{aa}(\vec r_0, t).
\label{Bloch_TLA}
\end{equation}
In the small-excitation limit, the term $\Gamma\rho_{aa}$
is of second order and can be neglected.
We thus arrive at the {\it linear} integral equation for the
atomic excitation
\begin{eqnarray}
\lefteqn{\dot\rho_{aa}(\vec r_0,t) \;=} \label{Holstein} \\
&& -\gamma\, \rho_{aa}(\vec r_0,t) + 
\gamma\, \int_V\!\! d^3\vec r\enspace
G(\vec r_0,\vec r)\, \rho_{aa}(\vec r,t).\nonumber \\
\end{eqnarray}
 Eq.(\ref{Holstein}) is the
integro-differential equation for radiation trapping
derived by Holstein in \cite{Holstein} for the special 
case of Doppler-broadened
two-level atoms. Thus in the linear limit we have rederived the
theory of radiation trapping of \cite{Holstein}. 


\subsubsection{dynamics of initially inverted two-level system
                in small-sample approximation}


Let us now discuss a nonlinear problem, but in a small volume, such that 
the space dependence can be disregarded. In this case we can carry out
the volume integral placing the probe atom on the axis of the long cylindrical
sample (see Fig.5).
 We find for the
decay rate for a probe atom with transition frequency $\omega$
\begin{equation}
\frac{\Gamma(\omega,t)}{\gamma}=
 \frac{\rho_{aa}(t)}{\rho_{bb}(t)-\rho_{aa}(t)}
\left[ 1-\exp\left(-K(t) e^{-\Delta^2/2\Delta_D^2}\right)\right],
\end{equation}
where $\Delta=\omega-\omega_{ab}^0$, and 
\begin{equation}
K(t)=K_0 d\, [\rho_{bb}(t)-\rho_{aa}(t)].
\end{equation}

Averaging over the inhomogeneous velocity distribution of the atoms
yields
\begin{eqnarray}
\lefteqn{\frac{\Gamma(t)}{\gamma}\;=\;
\frac{\rho_{aa}(t)}{\rho_{bb}(t)-\rho_{aa}(t)}} \nonumber \\
&& \frac{1}{\sqrt{\pi}}\, \int_{-\infty}^\infty\!\!\! dy\, e^{-y^2}
\left[ 1-\exp\left(-K(t) e^{-y^2}\right)\right].\label{G_TLA_approx}
\end{eqnarray}
Note that $\Gamma(t)/\gamma$ 
remains finite at $\rho_{aa}=1/2$,
since the diverging denominator is multiplied by a vanishing integral 
expression. 

The time evolution of the excited-state population from an initially
completely inverted system is shown in Fig.~6 for different values of 
the density $\eta\equiv N\lambda^2 d=100$ 
(solid line) and $\eta=500$ (dashed line)
and $g=0.01$.
The dotted line corresponds to the free-space decay. One recognizes 
a non-exponential behavior, with an accelerated
decay in the initial phase corresponding to amplified spontaneous 
emission and
a substantial slow-down of  decay in the long-time limit. 

\begin{figure}
\epsfxsize=6cm
\centerline{\leavevmode\epsffile{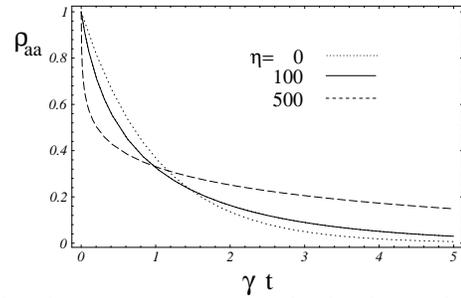}}
\caption{Time dependence of excitation in spatially homogeneous, dense
two-level medium. Time is in units of the inverse free-space decay rate.
$\eta\equiv N\lambda^2 d=0$ (dotted), $100$ (line), $500$ (dashed). 
$g=\gamma/\sqrt{2\pi}\Delta_D=0.01$.}
\end{figure}

The effective rate of decay of the excitation
$\Gamma_{\rm eff} = -{\dot\rho}_{aa}/\rho_{aa}$ is shown in Fig.~7.
One can see that for $\eta=500$ the initial decay rate is already 
of the order of the inhomogeneous Doppler-width 
(${\rm log}\Delta_D/\gamma \approx 1.6$) and the Markov-approximation
of slow atomic evolution becomes invalid. For higher atomic densities
the  system would show superradiant decay in the initial phase
which cannot be described by  the single-atom density matrix equation.
As noted before, modeling of the cooperative decay requires
a two-atom density matrix description, which will be discussed elsewhere
\cite{super}.
\begin{figure}
\epsfxsize=6cm
\centerline{\leavevmode\epsffile{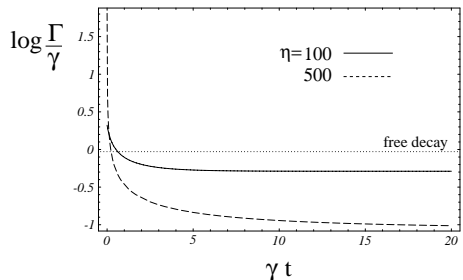}}
\caption{Effective rate of decay 
$\Gamma_{\rm eff}=-{\dot\rho}_{aa}/\rho_{aa}$
for examples of Fig.6. Dotted line
corresponds to free-space decay. Amplified spontaneous emission in the initial
phase and slow-down of decay in final phase are apparent.}
\end{figure}
One also verifies from Fig.~7 for the case $\eta=100$, that  the 
decay becomes exponential again in the long-time
limit.
The asymptotic escape rate 
is given by
\begin{equation}
\gamma_{\rm esc} = \frac{\gamma}{K_0(\pi \ln K_0)^{1/2}},
\end{equation}
which can be orders of magnitude smaller than $\gamma$.
This result agrees with  Eq.(1.1) of \cite{Holstein}b 
up to a numerical factor of the order of unity, which is due to the fact that
we here have disregarded a possible spatial inhomogeneity.

It is also instructive to consider the time-dependent 
spectrum of incoherent radiation or equivalently $\Gamma(\omega,t)$.
This is done in Fig.~8 for $\eta=500$. Shown is the spectral distribution
at different times normalized to the averaged rate $\Gamma(t)$. 
\begin{figure}
\epsfxsize=6cm
\centerline{\leavevmode\epsffile{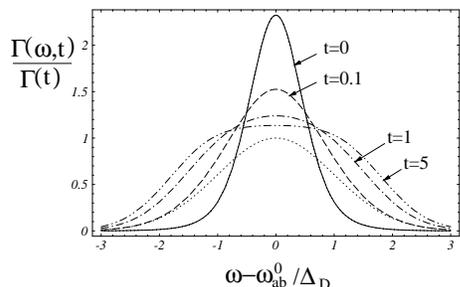}}
\caption{Spectral distribution of incoherent decay as function of time.
$\eta=500$, $g=0.01$. Dotted line shows Doppler-distribution  of atomic
transition frequencies (not normalized). Time is in units of $\gamma^{-1}$.}
\end{figure}
The dotted line shows the 
(not normalized) inhomogeneous distribution of atomic frequencies
according to Eq.(\ref{W}). One recognizes that the incoherent spectrum
broadens with the decay of excitation. In the initial phase of amplified 
spontaneous emission $(\gamma t=0\dots 1)$ 
one can see that the radiation spectrum
is narrower than the inhomogeneous atomic spectrum. This gives the first
indication of spectral condensation, a well-known phenomenon 
in amplifying media and lasers.


\subsection{Effects of radiative atom-atom interactions
            on intrinsic optical bistability}


One of the most interesting dynamical effects in dense media
due to the Lorentz-Lorenz nonlinearity is the possibility of intrinsic
optical bistability predicted in \cite{intr_bist}. If a radiatively 
broadened two-level system is resonantly driven by a coherent field
of Rabi-frequency $\Omega$ it shows mirrorless, i.e.~intrinsic bistability,
if the atomic density exceeds some critical value. The bistability results
from an effective feedback introduced by the Lorentz-Lorenz correction.

We here consider a dense ensemble of resonantly driven two-level systems
as shown in Fig.~9. For simplicity of the present discussion we
 assume that the driving field is homogeneous.
\begin{figure}
\epsfxsize=6cm
\centerline{\leavevmode\epsffile{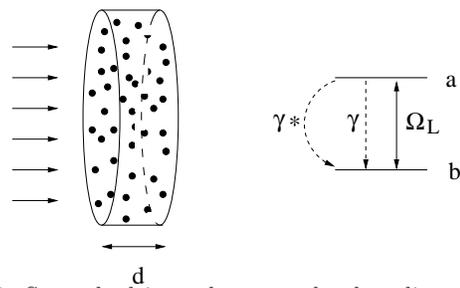}}
\caption{Strongly driven, dense two-level medium.
$\gamma$ and $\gamma^*$ describe radiative and non-radiative decays
respectively, and $\Omega_L$ denotes Rabi-frequency of local field.}
\end{figure}
For most practical realizations this assumption is not valid. We are here
however interested in principle questions and will therefore ignore
drive-field depletion. The assumption of a homogeneous driving field implies
a homogeneous behavior of the atomic system and we can disregard
the spatial dependence in the collective decay and light-shift terms.
The density matrix equations for the system under consideration
 read in a rotating frame
\begin{eqnarray}
\dot\rho_{aa} &=& - \Gamma_a\rho_{aa}+\Gamma \rho_{bb} 
- i(\Omega \rho_{ab} - \Omega
\rho_{ab}^*),\label{rho_aa}\\
\dot\rho_{ab} &=& -\Gamma_{ab}\rho_{ab} - i(\Omega+C\gamma\rho_{ab})
(\rho_{aa}-\rho_{bb}),\label{rho_ab}
\end{eqnarray}
where we have assumed a real $\Omega$.
$\Gamma_a=\gamma+\gamma^*+\Gamma$ 
is the total population decay rate out of the excited state,
with $\gamma$ and $\gamma^*$ being the free-space radiative and non-radiative
decay rates, and $\Gamma$ the collective decay rate. $\Gamma_{ab}=\Gamma+
(\gamma+\gamma^*)/2$. There is no collective light-shift contribution here
due to symmetry reasons. One recognizes a term proportional to the atomic
polarization $\rho_{ab}$ that adds to the Rabi-frequency $\Omega$.
This term is due to the Lorentz-Lorenz correction (\ref{Lorentz})
and has the character of a feedback (atomic polarization
generates a field contribution $C\gamma\rho_{ab}$ which 
couples back to the atom).
$C=N \lambda^3/4\pi^2$ is the cooperativity parameter, that essentially
determines the number of atoms in a volume $\lambda^3$. 

The stationary solution
 of Eqs.(\ref{rho_aa}) and (\ref{rho_ab}) for the excited state population 
for $\Gamma=\gamma^*=0$ is shown in Fig.~10 for different cooperativities.
One recognizes bistability for $C\ge 3$. 

\begin{figure}
\epsfxsize=6cm
\centerline{\leavevmode\epsffile{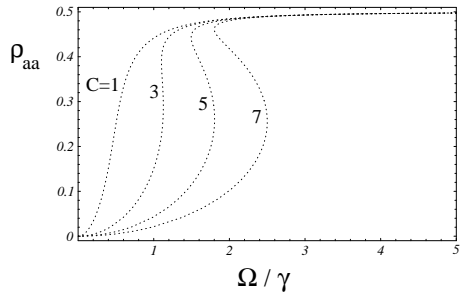}}
\caption{Stationary excited state population as function of 
driving-field Rabi-frequency $\Omega$ for different cooperativity
parameters. Here $\Gamma=\gamma^*=0$.}
\end{figure}

We now analyze the effect of 
incoherent photons inside the medium. 
To self-consistently determine the collective decay rate, we have to calculate
correlations of the dipole operators $\sigma=|b\rangle\langle a|$
 and $\sigma^\dagger$ in terms of density matrix elements. 
More precisely we need the second order cumulants,
i.e.~correlations of operators minus their mean values (which are nonzero
in the present case). Thus we 
start with the Heisenberg-Langevin equations for $\delta\sigma \equiv
\sigma-\langle\sigma\rangle$ and $\delta\sigma_{aa}=\sigma_{aa}-
\langle\sigma_{aa}\rangle$, where $\sigma_{aa}=|a\rangle\langle a|$:
\begin{eqnarray}
\delta{\dot\sigma} 
&=& -\Gamma_{ab}\delta\sigma - 2 i\Omega_L \delta\sigma_{aa}+
\, {\rm noise}\\
\delta{\dot\sigma}_{aa} &=& -2\Gamma_{ab}\delta\sigma_{aa}
-i \Bigl(\Omega_L^*\delta\sigma-\Omega_L\delta\sigma^\dagger\Bigr)
+\, {\rm noise},
\end{eqnarray}
where $\Omega_L=\Omega+C\gamma\rho_{ab}$.
The relevant correlations can be obtained from these equations by 
Laplace-transformation. This  yields at resonance $(\omega=\omega_{ab})$
\begin{eqnarray}
{\widetilde P}^{\rm ret}(\omega_{ab}) &=& \frac{\wp^2 N}{\hbar^2}
\frac{\Gamma_{ab}\bigl(\rho_{aa}-\rho_{bb}\bigr)}
{\Gamma_{ab}^2 + 2|\Omega_L|^2},\label{bistab_P_ret}\\
{\widetilde P}^{\rm\, s}(\omega_{ab}) &=& \frac{2\wp^2 N}{\hbar^2}
\frac{\Gamma_{ab}^2\Bigl(\rho_{aa}-|\rho_{ab}|^2\Bigr) 
+ 2|\Omega_L|^2 \rho_{aa}\rho_{bb}}
{\Gamma_{ab}\Bigl(\Gamma_{ab}^2 + 2|\Omega_L|^2\Bigr)}.\label{bistab_P_s}
\end{eqnarray}
Note that we have omitted space and time arguments, since
we are interested in the stationary properties of the system in
a homogeneous sample. 
From (\ref{bistab_P_ret}) we immediately find the absorption coefficient
\begin{equation}
q_0^{\prime\prime} =\frac{\wp^2 N}{3\hbar\epsilon_0} 
\frac{\omega_{ab}}{c}
\frac{\Gamma_{ab}}{\Gamma_{ab}^2 + 2 |\Omega_L|^2} \, \bigl(\rho_{aa}-\rho_{bb}
\bigr)\label{q_bistab}
\end{equation}

We now consider a thin plate of thickness $d$ as shown in Fig.~9 and assume
that the beam diameter of the driving field is large compared to $d$. 
Carrying out the spatial integrations in Eq.(\ref{G_sol}) using
(\ref{q_bistab}) and (\ref{bistab_P_s}) we find the following relation
for the collective decay rate
\begin{eqnarray}
\lefteqn{\frac{\Gamma}{\gamma} \;=} 
\label{bistab_G} \\
&&  \frac{1}{\rho_{bb}-\rho_{aa}}
\biggl(\rho_{aa} -|\rho_{ab}|^2 + \frac{2|\Omega_L|^2}{\Gamma_{ab}^2}
\rho_{aa}\rho_{bb}\biggr)\biggl(1-e^{2
q_0^{\prime\prime}d}\biggr). \nonumber 
\end{eqnarray}
This equation for $\Gamma$ is not yet explicit since
$\Gamma$ enters the right hand side of the equation in an essentially
nonlinear way.
Using the stationary solutions of the density matrix equations
(\ref{rho_aa}) and (\ref{rho_ab}) with $\Gamma$ as independent variable,
one can (with some additional approximations) 
solve  Eqs.(\ref{bistab_G}). This yields 
\begin{equation}
\frac{\Gamma}{\gamma}\approx \frac{\rho_{aa}}{1-2\rho_{aa}} 
\biggl(1-e^{-K}\biggr),\label{bistab_G_expl}
\end{equation}
with
\begin{equation}
K= C r\, \frac{1 + \gamma^*/\gamma (1- 2 \rho_{aa})}{
2\frac{\Omega^2}{\gamma^2} + 2 C^2 (1-2\rho_{aa})^2 +
\frac{(1+\gamma^*/\gamma(1-2 \rho_{aa}))^2}{4(1-2\rho_{aa})^2}}.
\end{equation}
$C$ is the cooperativity parameter, and $r=\pi d/\lambda$.

In Fig.~11 we show the stationary solutions for the excited state
population as function of the driving-field Rabi-frequency $\Omega$
for different cooperativity parameters and for purely radiative decay,
i.e. $\gamma^*=0$. The dotted curves correspond to the solutions without
radiative atom-atom interactions.
As can be seen, bistability persists, but cannot be 
resolved for physically reasonable values of $r$ ($r=100$ in Fig.~11b).
Radiation trapping
prevents the energy to escape from the sample and already
very small external pumping is sufficient to keep the atoms in a highly
excited state. 

\begin{figure}
\epsfxsize=6cm
\centerline{\leavevmode\epsffile{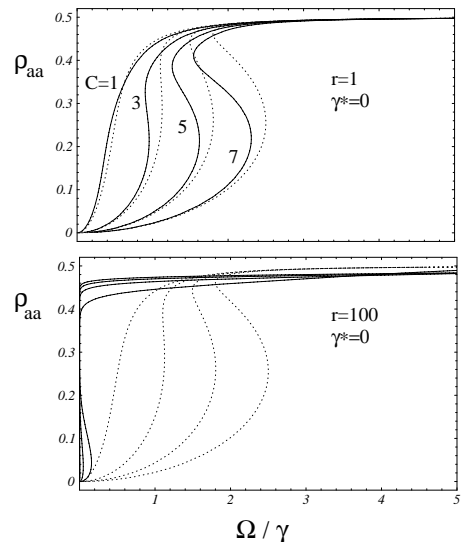}}
\caption{Stationary excited state population as function of 
driving-field Rabi-frequency $\Omega$ for
radiative decay ($\gamma^*=0$) and with radiative 
atom-atom interactions. $r= \pi d/\lambda=1$ (upper picture)
and $r= 100$ (lower picture). Dotted curves show behavior without
radiative atom-atom interactions (see Fig.~10.)}
\end{figure}

The situation is different, if there is also non-radiative decay, as shown 
in Fig.~12 (here $\gamma^*=\gamma$). In comparison to the 
radiatively-broadened
case of Fig.~11, the bistability curves are only 
moderately altered even for large samples with $r=1000$. The critical
cooperativity at which bistability starts to occur is somewhat increased. 
The non-radiative decay provides an 
additional energy escape channel, such that the trapped 
incoherent radiation 
is not strong enough  to keep the atoms in the excited state.   

Thus we can conclude, that radiative atom-atom interactions do not destroy
intrinsic bistability in driven two-level systems, if non-radiative
decay is present. This is different from our previous result 
\cite{Yelin97}. The reason for this discrepancy is,  that our
previous approach essentially neglected the medium effect on the
retarded propagation and was therefore inconsistent for larger 
densities. This would correspond to replacing the 
retarded propagator of the {\it interacting} field 
in Eq.(\ref{GF_eq1}) and Fig.~3 by the 
corresponding free-space Greensfunction. 

\begin{figure}
\epsfxsize=6cm
\centerline{\leavevmode\epsffile{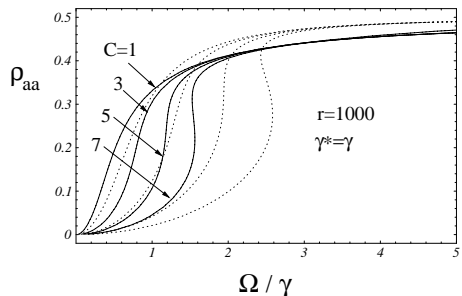}}
\caption{Stationary excited state population as function of 
driving-field Rabi-frequency $\Omega$ in the presence of
additional nonradiative decay ($\gamma^*=\gamma$) and with radiative 
atom-atom interactions. $r=1000$. Dotted curves show 
corresponding behavior without
radiative atom-atom interactions for $\gamma^*=\gamma$.}
\end{figure}

We note that we did not intend to present a comprehensive
discussion of effects that could affect intrinsic optical bistability.
In particular in atomic vapors collisions may have a much more pronounced
effect.
Furthermore the depletion of the pump field needs to be
taken into account and realistic experimental schemes such as
selective reflection spectroscopy \cite{refl_spectr} need to be considered.


\section{summary}


We have shown that the interaction of a classical radiation field
with a dense ensemble of atoms can be described by modified
Maxwell-Bloch equations in the Markov-limit. 
While the equations for the macroscopic
classical field, the Maxwell equations, remain unchanged, the
atomic equations of motion have additional nonlinear and spatially
nonlocal terms that result from the exchange of spontaneous photons
between the atoms. The first nonlinear term, the LL-correction, 
is only present if there is initial coherence
or an external coherent field. The nonlinear and nonlocal
collective decay  and level-shift terms are present whenever
there is excited-state population. 
In the Markov-limit of spectrally broad incoherent radiation inside the
medium, the modified Bloch equations have the form of single-atom 
density matrix equations. They are of the Lindblad type and thus
fulfill the formal requirements for conservation of probability and
positivity. The effect of the collective decay has been discussed
for the examples of an initially excited system of inhomogeneously broadened 
two-level atoms and intrinsic optical bistability. In the first case we
find accelerated decay (superluminescence) in the initial phase and 
radiation trapping in the final phase. In the long-time limit, where
the excited state population is small, the equations of motion 
coincide with the Holstein equations of radiation trapping \cite{Holstein}.
The collective decay modifies intrinsic optical bistability in a strongly
driven two-level system. As opposed to our previous prediction \cite{Yelin97}, 
bistability persists however, if also non-radiative decay is present.
Collective decay and pump processes as well
as light shifts are relevant for the population dynamics and are
particularly important for ground state coherences in multi-level
systems. A detailed discussion of coherence effects in dense 
multilevel systems as well as the study of cooperative decay processes
will be the subject of future work.


\section*{Acknowledgment}


The authors would like to thank G.~Agarwal, C.~M.~Bowden, J.~P.~Dowling,
M.~Lukin, A.~Manka, A.~Schenzle and M.~O.~Scully for stimulating
discussions. The support of the Office of Naval Research and the
Texas Advanced Research and Technology program are gratefully acknowledged.
M.F. would also like to thank the Alexander-von-Humboldt foundation
 and S.F.Y. the Studienstiftung des Deutschen Volkes and the German
Academic Exchange Service (through their program HSP II) for financial
support. 


\subsection*{Appendix A: Free-space decay rate}


In Eq.(\ref{Einstein_A}) we have given the spontaneous decay rate
in the atomic medium. We now show that this expression leads to the well-known
Wigner-Weisskopf result for the radiative decay of a two-level atom
if the interacting field is replaced by the free field.
For a single two-level transition with dipole moment along $\vec e_\mu$
we have according to (\ref{Einstein_A})
\begin{equation}
\gamma=\frac{\wp^2}{\hbar^2}\int_{-\infty}^\infty\!\!\! 
d\tau \Bigl\langle [ E_{0\, \mu}^+(\vec r_0,t+\tau),
E_{0\, \mu}^-(\vec r_0,t)]\Bigr\rangle
\, e^{i\omega\tau}.\label{gamma0}
\end{equation}
The free-field commutator is given by \cite{Pauli}
\begin{eqnarray}
\lefteqn{[ E_{0\, \mu}^+(\vec r_0,t+\tau),
E_{0\, \mu}^-(\vec r_0,t)]=} 
\label{commutator} \\
&&  \frac{\hbar c}{2\epsilon_0}
\frac{1}{(2\pi)^3}
\int d^3\vec k\, k \left(1- \frac{k_\mu^2}{k^2}\right)\, e^{-i ck\tau}.
\nonumber
\end{eqnarray}
Substituting (\ref{commutator}) into (\ref{gamma0})
yields
\begin{equation}
\gamma=\frac{\wp^2}{\hbar^2}\frac{\hbar \omega^3}{2\epsilon_0 c^3}
\frac{1}{(2\pi)^2} \int d^2\Omega_k\,  
\left(1- \frac{k_\mu^2}{k^2}\right)
\end{equation}
Finally carrying out the  angle-integration leads to
\begin{equation}
\gamma= \frac{\wp^2 \omega^3}{3\pi \hbar\epsilon_0 c^3} =\frac{8\pi^2 \wp^2}{
3\hbar\epsilon_0\lambda^3},
\end{equation}
which is the free-space spontaneous emission rate from the
Wigner-Weisskopf theory \cite{Louisell}.


\subsection*{Appendix B: Dyson equation for real-time Greensfunctions}


In this Appendix we derive the integral equations (\ref{GF_eq1})
and (\ref{GF_eq2}) for the {\it real-time} GF from the
Dyson equation (\ref{Dyson}) for the {\it contour} GF.
 Noting that $D_0^{-+}=0$ in RWA, we immediately
find from (\ref{Dyson})
\begin{eqnarray}
D^{++} &=& D_0^{++}
-   D_0^{++}\, \Pi^{++}\, D^{++}
+  D_0^{++}\, \Pi^{+-}\, D^{-+},
\label{A_D++}\\
&& \qquad\enspace
- D_0^{+-}\, \Pi^{--}\, D^{-+}
+  D_0^{+-}\, \Pi^{-+}\, D^{++},\nonumber\\
D^{-+} &=& \qquad\enspace - D_0^{--}\, \Pi^{--}\,
D^{-+} +  D_0^{--}\, \Pi^{-+}\,
D^{++},\label{A_D-+}
\end{eqnarray}
where we used a short notation
$ D_0^{--}\, \Pi^{--}\,
D^{-+} = \int\!\!\int d3 d4\,
 D_{0\,\alpha\mu}^{--}(1,3)\, \Pi^{--}_{\mu\nu}(3,4)\, D^{-+}_{\nu\beta}(4,2)$
and the integration goes over physical times from $-\infty$ to $\infty$
and over the volume of the sample.
Note the sign changes in Eqs.(\ref{A_D-+}) and (\ref{A_D++}) resulting
 from the fact 
that the contour integration on the lower branch goes
in the reverse direction.
Making use of (\ref{D0++RWA}-\ref{D0--RWA}) we obtain
\begin{eqnarray}
\lefteqn{D^{++} = D_0^{\rm adv} - D_0^{\rm
adv}\Bigl[\Bigl(\Pi^{++}-\Pi^{-+}\Bigr)  
D^{++}-}  \nonumber \\
&&\Bigl(\Pi^{+-} -\Pi^{--}\Bigr)D^{-+}\Bigr]
\label{A_D++2} - D_0^{\rm ret}\Bigl[
\Pi^{-+}\, D^{++} - \Pi^{--}\, D^{-+}\Bigr],\\
\lefteqn{D^{-+} = - D_0^{\rm ret}\Bigl[ \Pi^{-+}\, D^{++}
-\Pi^{--}\, D^{-+}\Bigr].} \label{A_D-+2}
\end{eqnarray}
Applying the relation 
\begin{eqnarray}
\Pi^{\rm adv}(1,2)
&\equiv& \Pi^{++}(1,2)-\Pi^{-+}(1,2)=\\
&& \Pi^{+-}(1,2)-\Pi^{--}(1,2)\nonumber 
\end{eqnarray}
and subtracting Eqs.(\ref{A_D-+2}) from  (\ref{A_D++2})
yields the Dyson equation for $D^{\rm adv}=D^{++}-D^{-+}$:
\begin{eqnarray}
\lefteqn{D^{\rm adv}_{\alpha\beta}(1,2)
=D_{0\,\alpha\beta}^{\rm adv}(1,2) -}\\
&& \int\!\!\int d3\, d4\, 
D_{0\,\alpha\mu}^{\rm adv}(1,3)\, \Pi^{\rm adv}_{\mu\nu}(3,4)\,
D^{\rm adv}_{\nu\beta}(4,2) \nonumber 
\end{eqnarray}
where we have restored full notation. Since $D^{\rm ret}_{\mu\nu}(1,2)
=D^{\rm adv}_{\nu\mu}(2,1)$ one immediately obtains the corresponding
Dyson equation for the retarded propagator 
\begin{eqnarray}
\lefteqn{D^{\rm ret}_{\alpha\beta}(1,2)
=D_{0\, \alpha\beta}^{\rm ret}(1,2) - } \\
&& \int\!\!\int d3\, d4\, 
D_{0\,\alpha\mu}^{\rm ret}(1,3)\, \Pi^{\rm ret}_{\mu\nu}(3,4)\,
D^{\rm ret}_{\nu\beta}(4,2) \nonumber
\end{eqnarray}
with
\begin{eqnarray}
\lefteqn{\Pi^{\rm ret}_{\mu\nu}(1,2)=\Pi^{++}_{\mu\nu}(1,2)-
\Pi^{+-}_{\mu\nu}(1,2)=} \nonumber \\
&& \Pi^{-+}_{\mu\nu}(1,2)-\Pi^{--}_{\mu\nu}(1,2) =\\
&& \frac{\wp_\mu\wp_\nu}{\hbar^2}\Theta(t_1-t_2)\nonumber \\
&& \qquad \sum_j\bigl\langle \bigl[ \sigma_{j\mu}^\dagger(t_1),
\sigma_{j\nu}(t_2)\bigr] 
\bigr\rangle\,\delta(\vec r_1\vec r_j)\, \delta(\vec r_2-\vec r_j).\nonumber
\end{eqnarray}
Thus we obtain the Dyson-equation (\ref{GF_eq2}) for the retarded Propagator
inside the medium of Sec.IV.

We now turn to $D^{-+}$. Substituting $\Pi^{--}=\Pi^{-+}-\Pi^{\rm ret}$
in (\ref{A_D-+2}) we find
\begin{equation}
D^{-+}=-D_0^{\rm ret} \Pi^{\rm ret} D^{-+} -
 D_0^{\rm ret} \Pi^{-+} D^{\rm adv}.
\end{equation}
Iteration of this equations yields
\begin{eqnarray}
D^{-+}&=&- \biggl[D_0^{\rm ret} -D_0^{\rm ret}\Pi^{\rm ret}
D_0^{\rm ret} + \\
&& D_0^{\rm ret}\Pi^{\rm ret} D_0^{\rm ret}\Pi^{\rm ret} D_0^{\rm ret}
-+\cdots\biggr]\,\Pi^{-+} D^{\rm adv} \nonumber
\end{eqnarray}
which can be rewritten in the compact form
\begin{equation}
D_{\alpha\beta}^{-+}(1,2) = -\int\!\!\int d3\, d4\, 
D_{\alpha\mu}^{\rm ret}(1,3)\, 
\Pi_{\mu\nu}^{\rm \, s}(3,4)\, D_{\nu\beta}^{\rm adv}(4,2),
\end{equation}
where
\begin{eqnarray}
\lefteqn{\Pi_{\mu\nu}^{\rm \, s}(1,2)\equiv\Pi_{\mu\nu}^{-+}(1,2)
=} \\
&& \frac{\wp_\mu\wp_\nu}{\hbar^2}\sum_j\Bigl\langle\Bigl\langle
\sigma_{j\mu}^\dagger(t_1)\sigma_{j\nu}(t_2)\Bigr\rangle\Bigr\rangle\,
\delta(\vec r_1-\vec r_j)\, \delta(\vec  r_2-\vec r_j). \nonumber 
\end{eqnarray}
Thus we arrive at Eq.(\ref{GF_eq1}) of Sec.IV.


\subsection*{Appendix C: Solution of the Dyson equation for 
${\widetilde D}^{\rm ret}$}


With the approximations made in Sec.IV-3 we derived the matrix solution 
(\ref{Dyson_matrix}) for the Fourier-transform 
of the retarded GF in the medium.
\begin{equation}
{\widetilde{\widetilde{\bf D}}}^{\rm ret}(\vec q,\omega,t)
=\left[{\bf 1} +\, {\widetilde{\widetilde{\bf D}}}^{\rm ret}_0(\vec q,\omega)
\cdot {\widetilde{\bf P}}^{\rm ret}(\omega,t)\right]^{-1}
\cdot {\widetilde{\widetilde{\bf D}}}^{\rm ret}_0(\vec q,\omega).
\label{A1}
\end{equation}
To evaluate this expression we first approximate the Fourier-transform of the
free-space retarded propagator. According to \cite{Pauli}
\begin{eqnarray}
&&D_{0\, \alpha\beta}^{\rm ret}(1,2)=\label{A2}\\
&&\quad\quad\frac{i\hbar}{4\pi\epsilon_0 c}\Theta(\tau)
\left[\delta_{\alpha\beta} \frac{\partial^2}{\partial \tau^2}-c^2\frac{
\partial^2}{\partial x_2^\alpha\partial x_2^\beta}\right]
\frac{\delta(r-c\tau)} {r},\nonumber
\end{eqnarray}
with $r=|\vec x|=|\vec r_1-\vec r_2|$ and $\tau=t_1-t_2$.
Thus
\begin{equation}
{\widetilde D}_{0\, \alpha\beta}^{\rm ret}(\vec x,\omega)
= -\frac{i\hbar}{4\pi\epsilon_0} \left(
\frac{\omega^2}{c^2}\delta_{\alpha\beta}
+\frac{x_\alpha x_\beta}{r^2}\frac{\partial^2}{\partial r^2}
\right)\, \frac{e^{-i\omega r/c}}{r}.
\end{equation}
For large $\omega$, such that $\lambda \ll r$, only the spatial derivative
of the exponential contributes and we find
\begin{equation}
{\widetilde D}_{0\, \alpha\beta}^{\rm ret}(\vec x,\omega)
= -\frac{i\hbar\omega^2}{4\pi\epsilon_0c^2} \left(
\delta_{\alpha\beta}
-\frac{x_\alpha x_\beta}{r^2}
\right)\, \frac{e^{-i\omega r/c}}{r}.\label{A3}
\end{equation}

We now approximate (\ref{A3}) by ignoring the
polarization, i.e.~by performing an orientation average.
\begin{equation}
\frac{x_\alpha x_\beta}{r^2} \longrightarrow \biggl\langle
\frac{x_\alpha x_\beta}{r^2} 
\biggr\rangle = \frac{1}{3}\, \delta_{\alpha\beta}.
\label{A4}
\end{equation}
This approximation is exact when the medium is randomly polarized.
(\ref{A4}) leads to
\begin{equation}
{\widetilde D}_{0}^{\rm ret}(\vec x,\omega)
 = -\frac{i\hbar\omega^2}{6\pi\epsilon_0 c^2} 
\frac{e^{-i\omega r/c}}{r}\label{A5}
\end{equation}
and thus
\begin{equation}
{\widetilde{\widetilde D}}_{0}^{\rm ret}(\vec q,\omega)
=-\frac{2 i\hbar \omega^2}{3 \epsilon_0 c^2}\, 
 \frac{1}{q^2-{\frac{\omega^2}{c^2}}
+ 2i\epsilon {\frac{\omega}{c}}},\label{A6}
\end{equation}
where we have introduced a small positive constant $\epsilon$ to move
the pole at $q=\omega/c$ into the lower half of the  complex plane.

Substituting (\ref{A6}) 
into (\ref{A1}) we  find
\begin{eqnarray}
{\widetilde D}^{\rm ret}(\vec x,\omega) &=& \frac{1}{(2\pi)^3}
\int d^3\vec q\enspace {\widetilde{\widetilde D}}^{\rm ret}(\vec q,\omega)
\, e^{-i\vec q\cdot\vec x}\label{A7}\\
&=& -\frac{\hbar\omega^2}{6\pi^2\epsilon_0 c^2} \frac{\delta_{\alpha\beta}}{r}
\int_{-\infty}^\infty\!\!\! dq \nonumber \\
&&\qquad\qquad \frac{q\, e^{-iq r}}
{q^2 -\frac{\omega^2}{c^2}\left[1 +
\frac{2 i\hbar }{3\epsilon_0}{\widetilde P}^{\rm ret}(\omega)
\right]+ 2i\epsilon\frac{\omega}{c}}.\nonumber
\end{eqnarray}
We evaluate this integral by contour integration along the real $q$-axis
and back in the lower half plane (note that $r>0$). 
In free space ($P=0$) and for $\epsilon=0$ we would have two poles
on the real axis at $q=\pm \omega/c$. In the free-space case we had
to introduce the constant $\epsilon>0$ to move the pole
at $q=\omega/c$ into the lower half-plane (and at the same time to
move the pole at $q=-\omega/c$ into the upper one). This is necessary to
have retarded propagation. For $\epsilon<0$ one would obtain the
advanced propagator. 
If an {\it absorbing} medium is present, ${\rm Re}[P^{\rm ret}]$ 
is negative and hence
adds to $\epsilon$. Therefore again the only contributing pole
is that at $q\approx \omega/c$. 
However, if the medium is {\it amplifying}
there is a problem. In this case ${\rm Re}[P^{\rm ret}]$ 
is positive and counteracts $\epsilon$. In such a case the pole
at $q=-\omega/c$ could move into the lower half-plane 
and we would obtain an advanced instead of a retarded propagator. 
The origin of this problem is, that the retarded propagator
in an amplifying medium is strictly speaking not Fourier-transformable, since
it is an exponentially growing function of distance. 
In such a case one has to take into account the finite spatial 
dimensions of the amplifying medium and 
introduce a cut-off function which leads to a
finite-space Fourier-transform.
Although this lacks mathematical rigor, we
now assume that the effect of the cut-off function is modeled
by a sufficiently large value of 
$\epsilon$, such that the pole at $q\approx \omega/c$
remains in the lower half-plane. With this we find 
\begin{equation}
{\widetilde D}^{\rm ret}(\vec x,\omega) =
-\frac{i\hbar \omega^2}{6\pi \epsilon_0 c^2}\frac{e^{
q_0^{\prime\prime} r}}{r}\, e^{-i q_0^\prime r},\label{A9}
\end{equation}
where
\begin{equation}
q_0(\vec r,\omega,t)=\frac{\omega}{c}\left[1+ \frac{ i\hbar }{3\epsilon_0} 
{\widetilde P}^{\rm ret}
(\vec r,\omega,t)\right].\label{A8}
\end{equation}


\end{document}